\documentclass{jetpl}
\twocolumn

\usepackage{amsmath,amssymb,color,pstricks,bm,alltt}

\begin{document}

\lat

\title{Hidden Fermi surface in K$_x$Fe$_{2-y}$Se$_2$: LDA+DMFT study}

\rtitle{K$_x$Fe$_{2-y}$Se$_2$ hidden Fermi surface}

\sodtitle{Hidden Fermi surface in K$_x$Fe$_{2-y}$Se$_2$: LDA+DMFT study}

\author{I.\,A.\,Nekrasov\/\thanks{e-mail: nekrasov@iep.uran.ru},
N.\,S.\,Pavlov N.S.\/\thanks{e-mail: pavlov@iep.uran.ru}}

\rauthor{I.\,A.\,Nekrasov, N.\,S.\,Pavlov}

\sodauthor{Nekrasov, Pavlov}

\address{Institute of Electrophysics, Russian Academy of Sciences,
Ural Branch, Amundsen str. 106,  Ekaterinburg 620016, Russia}


\abstract{
In this paper we provide theoretical LDA+DMFT support of recent ARPES experimental observation of the so called hidden hole like band and corresponding hidden Fermi surface sheet near  $\Gamma$-point in the  K$_{0.62}$Fe$_{1.7}$Se$_2$ compound. To some extent this is a solution to the long-standing riddle of Fermi surface absence around  $\Gamma$-point in the K$_x$Fe$_{2-y}$Se$_2$ class of iron chalcogenide superconductors. In accordance with the experimental data Fermi surface was found near the $\Gamma$-point within LDA+DMFT calculations. Based on the LDA+DMFT analysis in this paper it is shown that the largest of the experimental Fermi surface sheets is actually formed by a hybrid Fe-3d($xy,xz,yz$) quasiparticle band. It is also shown that the Fermi surface is not a simple circle as DFT-LDA predicts, but has (according to the LDA+DMFT) a more complicated ``propeller''-like structure due to correlations and multiorbital nature of the K$_x$Fe$_{2-y}$Se$_2$ materials. While the smallest experimental Fermi surface around $\Gamma$-point is in some sense fictitious, since it is formed by the summation of the intensities of the spectral function associated with ``propeller'' loupes and is not connected to any of quasiparticle bands.
}


\maketitle

\section{Introduction}

Investigation of the superconductivity in recently discovered iron-based superconductors is one of the main trends in modern condensed matter physics(see reviews~\cite{Sad_08, Hoso_09, John, MazKor, Stew, Kord_12}). Some of iron chalcogenide superconductors  have qualitatively different electronic properties from other iron-based superconductors (e.g. iron pnictides) \cite{PvsC}. Among them, the K$_x$Fe$_{2-y}$Se$_2$ compound and the FeSe monolayer on the SrTiO$_3$ substrate (FeSe/STO) take quite a special place \cite{Nekrasov_FeSe,Nekrasov_FeSe2,Nekrasov_FeSe3,Sad_16}. Early days angular resolved photoemission spectra (ARPES) experiments showed that for these compounds there are absent or practically can not be resolved hole-like  Fermi surface sheets near the $\Gamma$-point of the Brillouin zone\cite{Sad_16,kfese_arpes1,kfese_arpes2,FeSe_STO_arpes14}. While for the iron pnictides and some iron chalcogenides (e.g. bulk FeSe) these hole-like Fermi surface sheets near the $\Gamma$-point were clearly observed by ARPES (see e.g. Ref.~\cite{Sad_08}). Later for the K$_x$Fe$_{2-y}$Se$_2$ materials  ARPES showed around $\Gamma$-point some halo-like feature \cite{kfese_arpes3}.  The absence of the hole-like Fermi surface sheets near the  $\Gamma$-point indicates that for K$_x$Fe$_{2-y}$Se$_2$ series there is no possibility of nesting between the hole sheets of the Fermi surface near the  $\Gamma$-point and electronic sheets near the X-point. Thus a spin-fluctuation mechanism of superconducting pairing (assumed for iron pnictides \cite{MazKor}) is not applicable here. 

Recently in the work \cite{KFeSe_arpes16}  ARPES observation of a hidden hole-like band approaching the Fermi level near the $\Gamma$-point for the K$_{0.62}$Fe$_{1.7}$Se$_2$ system was reported. Also in the work \cite{KFeSe_arpes16} on the basis of the ARPES data analysis there was proposed a presence of a hidden hole-like Fermi surface near the $\Gamma$-point. The authors of \cite{KFeSe_arpes16} provide some reasons why the Fermi surfaces near the $\Gamma$-point previously were not observed due to the geometry of the experiment.

In the works \cite{KFeSeLDADMFT1,KFeSeLDADMFT2}, we already reported  theoretical LDA+DMFT study of K$_x$Fe$_{2-y}$Se$_2$ material.
The LDA$'$ calculations~\cite{CLDA,CLDA_long} of KFe$_2$Se$_2$ compound were
performed using the Linearized Muffin-Tin Orbitals method (LMTO)~\cite{LMTO}.
The ideal KFe$_2$Se$_2$ compound has tetragonal structure with the space group
$I$4/$mmm$ and lattice parameters $a=3.9136$~\AA~and $c=14.0367$~\AA~ were used.~\cite{KFeSe_cryst}.
The crystallographic positions are the following: K(2a)
(0.0, 0.0, 0.0), Fe(4d) (0.0, 0.5, 0.25), Se(4e) (0.0, 0.5, z$_{Se}$) with
z$_{Se}$=0.3539~\cite{KFeSe_cryst}.
For the DMFT part of LDA+DMFT calculations we employed CT-QMC impurity
solver~\cite{ctqmc,triqs}.
To define DMFT lattice problem for KFe$_2$Se$_2$ compound we used the full LDA
Hamiltonian (i.e. without any orbitals downfolding or projecting) same as in
Refs.~\cite{KFeSeLDADMFT1,KFeSeLDADMFT2}. Also recently quite extended discussion
of the origin of the shallow and ``replica'' bands in FeSe layered superconductors
was presented by our group within several recent publications \cite{Nekrasov_FeSe,Nekrasov_FeSe2,Nekrasov_FeSe3}.

Here we present LDA+DMFT results for the chemical composition K$_{0.62}$Fe$_{1.7}$Se$_2$ according to the ARPES results of Ref.~\cite{KFeSe_arpes16}.
One should stress that the composition investigated here is somewhat different from recently published results of Refs.~\cite{Nekrasov_FeSe,Nekrasov_FeSe2,Nekrasov_FeSe3}.

The DMFT(CT-QMC) computations were done at reciprocal temperature $\beta=40$
($\sim$290 K) with about 10$^8$ Monte-Carlo sweeps.
Interaction parameters of Hubbard model were taken $U$=3.75 eV, $J$=0.56 eV~\cite{KFeSe_arpes_UJ}. We employed the self-consistent
fully-localized limit definition of the double-counting correction~\cite{CLDA_long}.
Thus computed values of Fe-3d occupancies and corresponding double-counting
energies are $E_{dc}=18.50$, $n_d=5.66$.

The LDA+DMFT spectral function maps were obtained after analytic continuation
of the local self-energy $\Sigma(\omega)$ from Matsubara frequencies to the real
ones. To this end we have applied Pade approximant algorithm~\cite{pade} and
checked the results with the maximum entropy method~\cite{ME} for Green's
function G($\tau$).

\begin{figure*}[!t]
	\center{\includegraphics[width=1.\linewidth]{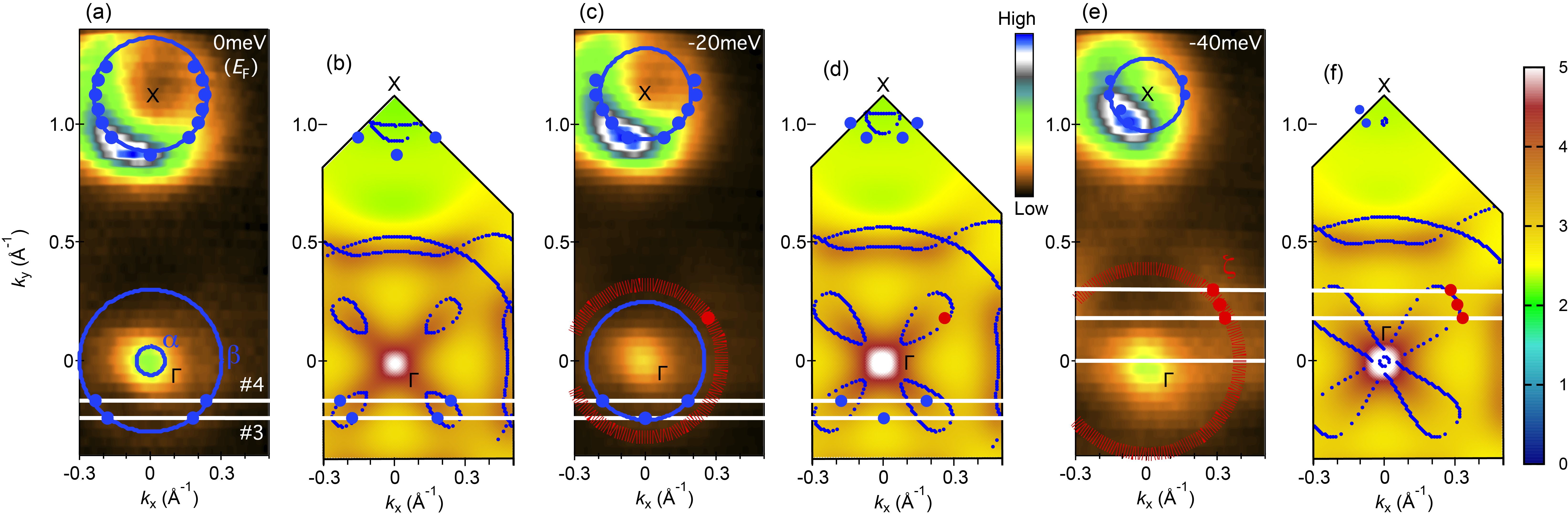}}
	\caption{Fig. 1. Panels -- (a),(c),(e) ARPES Fermi surface maps for K$_{0.62}$Fe$_{1.7}$Se$_2$~\cite{KFeSe_arpes16}
	plotted with different offset energies with respect to experimental Fermi level energy with Fermi surface sheets shown by dots and lines to guide eyes;
		Panels (b),(d),(f) -- LDA$'$+DMFT Fermi surface maps with maxima of corresponding spectral function shown by small blue dots.
		Big blue and red dots on (b),(d),(f) panels correspond to dots on (a),(c),(e) panels.}
	\label{fig1}
\end{figure*}

\section{Results and discussion}

 \begin{figure*}[!hbtp]
	\center{\includegraphics[width=1.\linewidth]{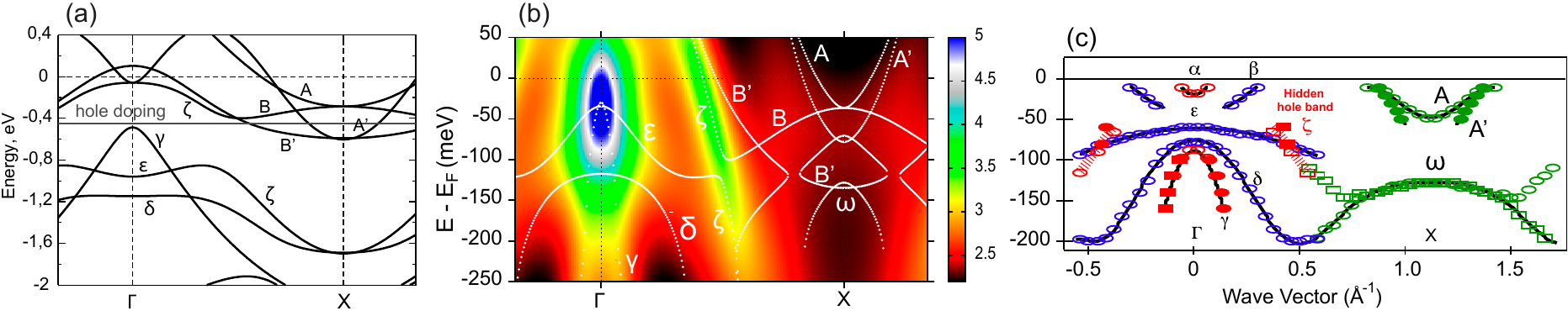}}
	\caption{Fig. 2. Panel (a) -- LDA$'$ band dispersions of paramagnetic KFe$_2$Se$_2$ (from Ref. \cite{Nekrasov_FeSe2});
	panel (b) -- LDA$'$+DMFT spectral function map with maxima shown by white crosses
		for K$_{0.62}$Fe$_{1.7}$Se$_2$;
	panel (c) -- quasiparticle bands extracted from
		ARPES~\cite{KFeSe_arpes16}.
		The letters designate same bands on different panels. The Fermi level is at zero energy.}
	\label{fig2}
\end{figure*}

In the Fig. 1 we present comparison of ARPES Fermi surface maps from Ref.~\cite{KFeSe_arpes16} (panels (a),(c),(e)) with our theoretical LDA$'$+DMFT
data (panels (b),(d),(f)). Experimental data \cite{KFeSe_arpes16} (panels (a),(c),(e)) is given for three different offset energies with respect to experimental Fermi level energy -- 0, -20~meV and -40~meV. On the panel (a) experimentalists see one Fermi surface sheet near X-point. For the Fermi surface sheet around X-point there are many experimental points shown by blue dots which finally form a circle-like sheet (drawn by blue line as a guide for an eye). The same situation is also seen on panels (c) and (e) with narrowing of the sheet upon going away from the Fermi level. This picture coincides with many other experimental papers (see e.g. review \cite{Sad_16}). A similar behavior near X-point can be noted for LDA+DMFT results on panels  (b),(d) and (f). However LDA+DMFT gives here two small tightly located Fermi sheets, which probably are not resolved by ARPES. These sheets are formed by A and A' bands of  Fe-3d($xy,xz,yz$) characters as can be viewed in Fig. 2 for LDA (panel a), LDA+DMFT (panel b) and experimental (panel c) results. This situation is similar to several earlier ARPES papers \cite{kfese_arpes1,kfese_arpes2,kfese_arpes3} for slightly different chemical compositions of   K$_x$Fe$_{2-y}$Se$_2$.

\begin{figure*}[!hbtp]
	\center{\includegraphics[width=.85\linewidth]{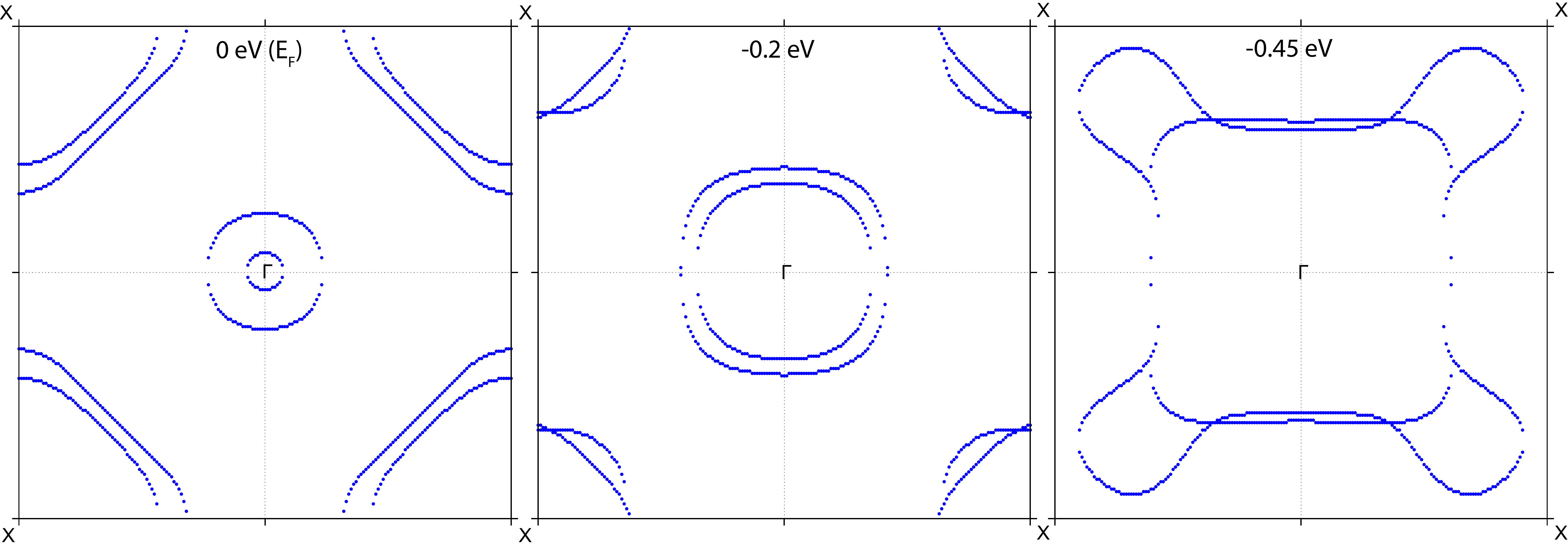}}
	\caption{Fig. 3. LDA calculated Fermi surface contours for KFe$_2$Se$_2$ 
	plotted with different offset energies with respect to LDA Fermi level energy (different hole doping).}
	\label{fig3}
\end{figure*}

The situation in the vicinity of the $\Gamma$-point is physically much more interesting. The authors of the Ref.~\cite{KFeSe_arpes16} claim that there are two Fermi surface sheets near $\Gamma$-point marked as $\alpha$ and $\beta$ (see Figs. 1(a) and 2(c)). Those $\alpha$ and $\beta$ sheets the authors of the Ref.~\cite{KFeSe_arpes16} call as a hidden Fermi surface. However for the Fermi surface sheet around $\Gamma$-point there are only four points for which expermintalists assume a circular shape of the sheet (drown by blue line to guide eyes) as a simplest possible choice. But if we look at panel (b) of Fig. 1 containing  LDA+DMFT data it immediately appears that around  $\Gamma$-point there are four quite small ``propeller'' like Fermi surface  sheets. Surprisingly,  if we map experimental blue dots from panel (a) to theoretical Fermi surface map (panel (b)) those dots perfectly coincides with ``propellers''. Thus one can conclude that indeed around $\Gamma$-point there is a manifestation of the Fermi surface with a more complicated shape in contrast to a circular one suggested in  Ref.~\cite{KFeSe_arpes16}. However to be honest similar Fermi surface sheets around $\Gamma$-point were resolved by ARPES before in the work \cite{kfese_arpes3}. Even more in the Ref. \cite{kfese_arpes3} in Fig. 1(c) one can see those ``propellers'' (not underlined by the authors) but for a bit different chemical composition  of   K$_x$Fe$_{2-y}$Se$_2$. 

If we now turn to the tiny $\alpha$ sheet on panel (a) of Fig. 1 one can observe rather a halo-like structure than a circle one. On the panel (b) of Fig. 1 one can find identical halo at the  $\Gamma$-point which is formed because of summation of intensities coming from the ``propeller'' blades. So the work of Ref.~\cite{KFeSe_arpes16} shows a presence of a hidden hole like Fermi surface near  $\Gamma$-point for KFeSe-class of systems. Similar behavior can be observed on the panels (c),(d) and (e),(f). 
 
 Remarkably, such ``propellers'' can not be obtained just from LDA data as can be seen in Fig. 3. Panels (a), (b) and (c) of Fig. 3 correspond to the same hole doping levels as shown in Fig. 1. For small doping (panels (a) and (b) of Fig. 3) there are circle-like structures near X-point similar to those on Fig. 1 (panels (a,c)).
Also one should say here that the  ``propellers'' are not a result of trivial renormalization of the bandwidth due to mass renormalization. Thus  ``propellers'' are essentially correlation effect plus multiorbital effects. Namely, real part of LDA+DMFT self-energy plays essential role to shift different bands with respect to each other. Strictly speaking LDA bands can form propellers in case Fermi level is about -0.9 eV, but with lobes in the 
$\Gamma$-X direction. While  LDA$'$+DMFT gives ``propellers'' turned to 45 degrees around $\Gamma$-point.

One should mentioned here that  LDA$'$+DMFT Fermi surfaces demonstrate another one sheet appearing in the middle of $\Gamma$-X direction. It can be seen on panels (b), (d) and (f) of Fig. 1. Somewhat similar motif could be observed on panel (c) of Fig. 3. This Fermi surface sheet is formed by Fe-3d($t_{2g}$ orbitals. However it is not resolved in the ARPES although hidden hole band is there on the panel (c) of Fig. 2. But the band disappears in the experiment and does not enter energies about -0.4 eV.

The situation with Fermi surface for KFeSe-class of materials as it is shown above is quite complicated. This differs intercalated iron chalcogenide superconductors  from iron pnictides. To illustrate that we present here comparison of LDA and LDA+DMFT Fermi surfaces for typical iron pnictide compound NaFeAs (not reported in our previous work \cite{NaFeAs}). In Fig. 4 one can see that in general LDA Fermi surface contours (on the left) are very much the same with LDA+DMFT Fermi surface map (on the right). Thus correlation effects as we mentioned in our previous works for iron arsenides do not influence Fermi surface significantly (see also Ref.~\cite{Sad_08}). While for chalcogenides correlation effects provide rather strong reconstruction of the quasiparticle bands in the close vicinity of the Fermi level \cite{KFeSeLDADMFT1,KFeSeLDADMFT2}.

\section{Conclusion}
To conclude, here on the basis of the works \cite{KFeSeLDADMFT1,KFeSeLDADMFT2} and inspired by the work of Ref.~\cite{KFeSe_arpes16} we confirm within our LDA+DMFT calculations that for K$_{0.62}$Fe$_{1.7}$Se$_2$ system near the $\Gamma$-point there are the hidden Fermi surface sheets. Also it is demonstrated that correlation effects are more important for KFeSe-supercoductors than for FeAs-based materials in a sense of Fermi surface formation. Possibly its appearance can partially justify spin-fluctuation mechanism of superconductivity in this class of systems with a rather high critical temperature T$_c\sim$30K. Good qualitative and even quantitative agreement of the calculated and ARPES Fermi surfaces is obtained.

\begin{figure}[!ht]
	\center{\includegraphics[width=.85\linewidth]{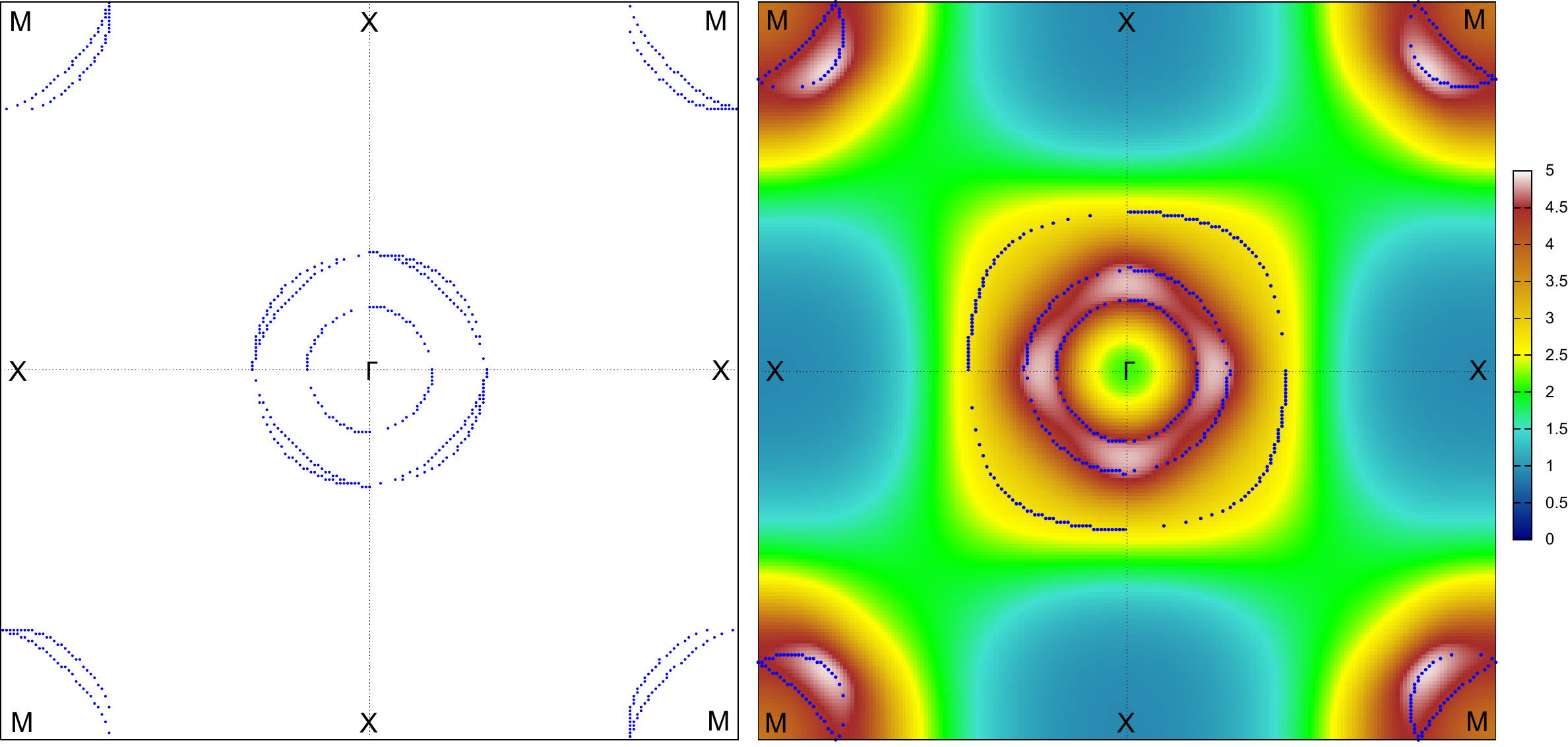}}
	\caption{Fig. 4. Left panel LDA calculated Fermi surface contours for NaFeAs; 
	right panel -- LDA+DMFT Fermi surface maps with maxima of corresponding spectral function shown by small blue dots.}
	\label{fig4}
\end{figure}

\section{Acknowledgments}
We thank M.V. Sadovskii and M.M. Korshunov for useful discussions.
This work was done under the State contract (FASO) No. 0389-2014-0001 and
supported in part by RFBR grant No. 17-02-00015.
Also we are grateful to the Program of Fundamental Research of RAS N12 ''Fundamental problems of high-temperature superconductivity'' for support.

\end{document}